\documentclass{article}
\usepackage{spconf,amsmath,epsfig}

\usepackage{graphicx,subfig}
\usepackage{common}
\usepackage{color}
\usepackage{url}
\usepackage{multirow} 
\usepackage{array}
\usepackage{numprint}

\title{Assessing the Quality of Wikipedia Pages Using Edit Longevity and Contributor Centrality}

\name{Xiangju Qin, P\'{a}draig Cunningham}
\address{School of Computer Science \& Informatics\\University College Dublin\\Email: xiangju.qin@ucdconnect.ie,padraig.cunningham@ucd.ie}

\begin{document}
%\ninept
%
\maketitle
\begin{abstract}
In this paper we address the challenge of assessing the quality of Wikipedia pages using scores derived from edit contribution and contributor authoritativeness measures. The hypothesis is that pages with significant contributions from authoritative contributors are likely to be high-quality pages. Contributions are quantified using edit longevity measures and contributor authoritativeness is scored using centrality metrics in either Wikipedia talk or co-author networks. The results suggest that it is useful to take into account the contributor authoritativeness when assessing the information quality of Wikipedia content. The percentile visualization of the quality scores provides some insights about anomalous articles, and could be used to help Wikipedia editors to identify Start and Stub articles that are of relatively good quality.
\end{abstract}
%

%!TEX root = AICS2012-Wikipedia.tex
\section{Introduction}

%Despite the widespread adoption of wikis and other social media, little is understood about how users might collaborate most effectively when using these tools.
%its open nature: anyone can contribute to its articles simply by clicking on an “edit” button.
% Nodes with high values of betweenness make the spreading of cutting-edge knowledge easier; in the case of Wikipedia it could be policies and best practices.
% the less experienced users <--> inexperienced users		%tend to interact with the less experienced users

In recent years, the world has witnessed an exponential growth of User-Generated Content (UGC) applications. Among these applications, Wikipedia is the most successful one with the aim of harnessing the contributions of millions of individuals to build a free collaborative encyclopedia. Wikipedia has attracted millions of visits everyday, and has become one of the most widely used sources of information on the web. On the other hand, Wikipedia articles are constantly changing, and the contributors range from casual visitors, to professionals and dedicated editors. When visitors access Wikipedia content linked through search engines, they are presented with the latest version of the article but they have no idea how much the content can be relied upon. Despite its great success as a means of knowledge sharing and collaboration, it is still difficult for visitors to develop an informed opinion about the reliability of much of the content available on Wikipedia. These issues have generated an increasing interest in studying the assessment of the trustworthiness of Wikipedia content \cite{Kane2011,adler2008b,adler2007,Hu2007}.

By extensive study on the co-author network (two contributors edited the same page establish a co-authorship) of the English Wikipedia community, Laniado and Tasso find that a nucleus of very active contributors, who seem to spread over the whole wiki, tend to interact preferentially with the less experienced users \cite{Laniado2011}. This finding is supported by the growing centrality of the very active contributors in the co-author network. These dedicated editors play a fundamental role in the community of Wikipedia in terms of spreading knowledge, information and experience across the whole wiki. In this paper, we explore the idea of assessing the quality of Wikipedia articles leveraging the scores derived from edit contributions and contributor authoritativeness metrics. Edit contributions are quantified using edit longevity measure and contributor authoritativeness is scored using network centrality metrics in either the Wikipedia talk or co-author networks. While the former captures author contributions recorded in the complete edit history of the articles, the latter measures contributor authoritativeness that encodes the communication patterns in the wikipedia networks. The intuition is that articles with significant contributions from authoritative contributors are likely to be of high quality, and that high-quality articles generally involve more communication and interaction between authors. By incorporating this information into the assessment of the quality of Wikipedia articles, we expect to develop a better strategy to assess the quality of Wikipedia content.

In the next section we provide a brief review of some relevant studies on the assessment of information quality for Wikipedia content and network analysis of Wikipedia. Then in Section 3 we introduce the edit longevity metric used to calculate the contribution of each author to a page, while in Section 4 we describe the centrality metrics that we use to measure contributor authoritativeness. Our models used to assess the quality of Wikipedia articles are presented in Section 5, and an evaluation of the models is provided in Section 6. The paper concludes in Section 7 with some discussions and an outline of future work.

%!TEX root = AICS2012-Wikipedia.tex
\section{Related work}\label{sec:related}

% four collaborative features (i.e., volume of contributor activity, type of contributor activity, number of anonymous contributors, and top contributor experience). The quantitative analysis suggests that certain features(e.g., content shaping, depth of top contributor experience) are positively associated with the quality of peer-produced information, while some (e.g., anonymous contributors, breadth of top contributor experience) have a negative influence on the information quality, and the total volume of contributions and contributors have no relationship with the quality of peer-produced information \cite{Kane2011}. 

Recently, researchers have shown an increased interest in measuring the quality or trustworthiness of UGC and Wikipedia content in particular. For instance, Adler {\it et al.} propose to make use of a trust quality metric (i.e., author reputation based on edit longevity) to measure the reliability of Wikipedia content \cite{adler2007,adler2008b}. Hu {\it et al.} propose several models to assess the quality of Wikipedia articles and contributor authority based on the assumption that “good contributors usually contribute good articles and good articles are contributed by good authors” \cite{Hu2007}. However they neglect the fact that people who have high authority and knowledge may only possess that for a specific domain. Moturu and Liu evaluate the trustworthiness of social media content using feature categories identified from sociological theory, and adopt unsupervised trust scoring models to combine these features \cite{Moturu2010}. Different from previous studies, Kane \cite{Kane2011} performs a quantitative study of the collaborative features associated with 188 similar high-quality Wikipedia articles in an attempt to better understand the mechanism behind the success of peer-produced collaboration in wiki environments. In the study, the author examines the relationship between the quality of Wikipedia articles and the four collaborative features (i.e., volume of contributor activity, type of contributor activity, number of anonymous contributors, and top contributor experience) associated with each article. The quantitative analysis suggests that different sets of features have different influence on the quality of peer-produced information. Later, Kane and Ransbotham \cite{Kane2012} study the relationship between collaboration patterns and the quality of Wikipedia articles by using ordinal regression. In their study, collaboration patterns are measured in terms of the number of distinct contributors to each article, degree centrality and eigenvector centrality of each article in the two-mode network. By experiment on 16,068 articles from the Medicine WikiProject, the authors show that there is a recursive, positive correlation between  quality and collaboration on Wikipedia articles \cite{Kane2012}. Wu {\it et al.} \cite{Wu2011} propose to characterize the quality of Wikipedia articles solely using network motif profiles, and demonstrate that the network motif-based characterization can be used to classify good from ordinary quality articles with reasonable accuracy. However, to our knowledge, very few researchers propose to assess the quality of Wikipedia articles by combining the influence of authors in the network and their edit contributions derived from the edit history of the articles.

%The article nodes are linked by hyperlinks and contributors are linked if they have worked on the same article. Contributors are also linked to articles on which they worked. The study proposes article and contributor degree centrality as indicators of authoritativeness. 

Korfiatis {\it et al.} \cite{Korfiatis2006} propose a network-based approach to evaluate authoritative sources in Wikipedia by using the centrality metrics from a two-mode network of articles and contributors. They evaluate their quality measure on a small dataset consisting of ten articles in the ``Philosophy'' domain from the English Wikipedia, and suggest that it could be useful to utilize the social network measures to evaluate the authoritativeness of content found in Wikipedia and similar sour-ces. This study is similar in spirit to the strategy in our work as centrality metrics are used to measure contributor authoritativeness.

There have been many quantitative studies on Wikipedia content to measure author contribution, such as the number of edits performed by authors (e.g., \cite{Wilkinson2007,Kittur2007}), the total number of words introduced by contributors \cite{Kittur2007}, text survial and edit distance \cite{adler2007,adler2008a}. Adler {\it et al.} propose a set of metrics and efficent algorithms to compute author contributions, and show that edit longevity is a good indicator of author contribution \cite{adler2008a}. While the two widely used criteria, edit count and text count, are naive and easy to compute, they fail to capture the size or the quality of the contributions. In contrast, edit longevity takes into account the amount of edits performed by an author and the survival of these edits in the subsequent revisions. In this work, we adopted the edit longevity metric \cite{adler2007,adler2008a} to measure author contributions to a wiki page, both for its accuracy and its efficient computation using the open source WikiTrust software\footnote{http://wikitrust.soe.ucsc.edu}.

Some researchers have studied the co-author network of Wikipedia, aiming at finding patterns of collaboration and cooperation in the process of Wikipedia content creation. These studies are mainly based on the simple assumption that two contributors edited the same page is enough to establish a co-authorship (e.g., \cite{BiukAghai2006}), and usually fail to scale to the size of Wikipedia in the major languages such as English. Laniado and Tasso \cite{Laniado2011} utilize edit longevity to compute a score to evaluate the contribution of each contributor to each wiki page, then select the main contributors for the page according to the scores, the co-author network is then constructed based on the selected main contributors for each page. We employed this approach to generate the co-author network for the English Wikipedia.

To facilitate direct communication between Wikipedians, Mediawiki software assigns to every registered user a user page and a user talk page (i.e., UTP). Similar to Wikipedia articles, these pages can be edited by anyone. Any user can leave another user a message by editing their UTP, the owner can choose to reply to a message on her own UTP or on the desired receiver's UTP. Massa \cite{Massa2011} extracts the communication network for Venetian Wikipedia users by reading and coding the messages left on UTP conversations. Specifically, for each message written by user A on user B's talk page, the two users were added as nodes to the network and a corresponding edge from A to B was created, with the weight of the edge representing the number of messages A wrote to B. Analyzing the social network of Wikipedia may provide a deeper insight into the social dynamics of Wikipedia, and reveal communication patterns (i.e., the flow of knowledge, experience and Wikipedia rules) among Wikipedia users \cite{Massa2011}. We chose to rely on the open source wiki-network software\footnote{https://github.com/phauly/wiki-network} released by Massa to construct the talk network for English Wikipedia. This software provides two algorithms to generate the talk network from UTPs, i.e., signature2graph.py and utpedits2graph.py. The former generates the talk network by parsing and counting signatures on the current version of UTPs in the current data dump, the latter builds the talk network on the complete dump. Since the history algorithm extracts the talk network from the whole edit history of UTPs, it generally captures more communications and interactions among the users. As stated by the author in \cite{Massa2011}, since the existing signatures are not affected by a rename, the signature algorithm usually fails to detect the rename issue, while the history algorithm is not affected by this issue. In this work, we consider and compare the centrality metrics from the two talk networks in assessing the quality of Wikipedia articles.

%!TEX root = AICS2012-Wikipedia.tex
\section{Edit Longevity}

%Explain how edit longevity is calculated.
%\cite{adler2007,adler2008a}
%Explain how we use it in scoring.
% the lifespan of a word
 
In this work, we employ edit longevity, proposed by Adler {\it et al.} \cite{adler2008a}, to measure edit contribution of each contributor to a Wikipedia article. The computation of edit longevity can be summarized as follows \cite{adler2008a}: Suppose we have a wiki page $p$ with $n>0$ versions $v_{0}, v_{1}, v_{2}, ..., v_{n}$, the initial version $v_{0}$ is empty, the $i$-th version $v_{i}$ ($i \in [1,n]$) is obtained by an author editing a revision: $v_{i-1}\rightarrow v_{i}$. Since each revision is edited by only one author, we denote the author who performed revision $r_{i}$ as $a_{i}$. The edit contribution made in a revision $r_{i}$ is defined as $d(r_{i}) = d(v_{i-1},v_{i})$. The edit distance between two versions, $v_{i}$ and $v_{j}$, is computed by
\begin{equation}
   d(v_{i},v_{j})= max(I,D)-\frac{1}{2}min(I,D)+M%\nonumber
\label{formula:EditDistance}
\end{equation}
where $I(v_{i},v_{j})$ is the number of words that are inserted, \\$D(v_{i},v_{j})$ represents the number of words that are deleted, $M(v_{i},v_{j})$ denotes the number of words that are moved, times the fraction of the document that they move across. The details of this definition can be found in \cite{adler2007}. 

The edit distance between $v_{i}$ and $v_{j}$ is a quantity measure, and it evaluates how much change (measured in terms of word additions, deletions, replacements, disreplacements, etc.) there has been going on from $v_{i}$ to $v_{j}$. The quality (also termed as longevity) of the edits performed in revision $r_{i}$ of a page (corresponding to $v_{i}$) is defined as follows \cite{adler2007,adler2008a}:
\begin{equation}
\label{formula:EditQuality}
   \alpha_{edit}(v_{i},v_{j})= \frac{d(v_{i-1},v_{j})-d(v_{i},v_{j})}{d(v_{i-1},v_{i})} %\nonumber
\end{equation}
The focus here is on the quality of edit $r_i$ that brought the page from $v_{i-1}$ to $v_{i}$. The objective is to get an assessment of how much of that edit has survived in version $v_{j}$.
Adler {\it et al.} take special care to make sure that the edit distance $d(v_{i},v_{j})$ satisfies the triangular inequality, so that $\alpha_{edit}(v_{i},v_{j})$ takes values from -1 to +1. $\alpha_{edit}(v_{i},v_{j})$=-1 means the revisions performed by $a_{i}$ are completely reverted by the following authors, $\alpha_{edit}(v_{i},v_{j})$=+1 indicates the edit contribution made by $a_{i}$ in revision $r_{i}$ are completely preserved by other authors \cite{adler2007,adler2008a}. When the value of $\alpha_{edit}(v_{i},v_{j})$ does not fall into [-1,+1], we can trim it to one of these two values. 

Due to frequent vandalism that happens in the wiki, it is a good idea to judge the edit quality using several succeeding revisions. Let us denote $J_{r_{i}}$ as the set of the first ten versions after $r_{i}$ that have authors different from that of $r_{i}$. For $J_{r_{i}}\neq \phi$, the average edit quality $\overline {\alpha}_{edit}(v_{i},v_{j})$ of $r_{i}$ is defined as follows \cite{adler2008a}:
\begin{eqnarray}
\label{formula:AvgEditQuality}
   \overline {\alpha}_{edit}(v_{i},v_{j}) = \frac{1}{|J_{r_{i}}|}\cdot \left ( \sum_{r_{j} \in J_{r_{i}}} \alpha_{edit}(v_{i},v_{j}) \right )  %\nonumber
\end{eqnarray}

The edit longevity is computed by combining the size of the edit performed by an author and the longevity of the edit in the following revisions. Thus, the edit longevity of a revision $r$ made by an author $a$ can be defined as:
\begin{equation}
\label{formula:EditLongevity}
   EditLong(r) = \overline {\alpha}_{edit}(v_{i},v_{j}) \cdot d(r) %\nonumber
\end{equation}
Similar to the work in \cite{Laniado2011}, we are also only concerned with the positive contribution carried by each author to a page, and neglect revisions bringing a negative score. We denote $E_{a,p}$ as the set of edits performed by author $a$ on page $p$, then the contribution of author $a$ to page $p$ can be computed by accumulating the edit longevity over all revisions (i.e., edits) performed by this author as follows:
\begin{eqnarray}
\label{formula:AuthorContribution}
   contrib(a,p) = \sum_{e \in E_{a,p}| EditLong(e)>0} EditLong(e) %\nonumber
\end{eqnarray}

We select the main contributors for each page using a similar strategy to \cite{Laniado2011}: as the anonymous users in Wikipedia do not have reliable nicknames, we discard all anonymous contributions. Let us denote the set of all registered users who edited page $p$ as $U_{p}$, ordered by descending edit longevity. The set of authors for page $p$ are selected as the subset $A_{p} \subset U_{p}$ including the first users of $U_{p}$ that make at least $\theta$ percent of edit contribution to $p$, i.e., 
\begin{eqnarray}
\label{formula:AuthorSelection}
   \frac{\sum_{a \in A_{p}| contrib(a,p)>M, |A_{p}|\geq K } contrib(a,p)}{\sum_{u \in U_{p}} contrib(u,p)} > \theta    %\nonumber
\end{eqnarray}
where $M$ is the minimum threshold for author contribution, $K$ is the minimum number of authors considering as authors of $p$, $\theta \in [0,100]$ is the percentage contribution threshold. Considering that Wikipedia is the result of peer-produced collaboration, we impose the $|A_{p}|\geq K$ constraint on author selection so that we can select more authors even for those pages with their edit contributions dominated by a few contributors.

%!TEX root = AICS2012-Wikipedia.tex
\section{Author Centrality}

%Describe the Talk and Co-Author networks.
%Explain how the centrality scores are calculated.

In this work, we aim at measuring contributor authoritativeness using centrality measured in either the talk network and co-author network of the Wikipedia. In the 1950s, Bavelas introduced the idea of correlating the position of an individual in a social network with the relative influence or importance of that individual in the network for organizational communication \cite{Bavelas1950}. Since then, many centrality metrics (e.g., degree, betweenness, eigenvector, etc) have been proposed to investigate the influence and special properties of individuals in a network (e.g., \cite{Russo2005,Freeman1979}). In this work, we mainly adopt three widely-used certrality measures to assess contributor authoritativeness: degree, betweenness and eigenvector centrality.  

% degree centrality: index of communication activity
% betweenness : index of potential for control of communication
% closeness : index of independence or efficiency		The closeness of an individual in a network is a function of the inverse of the average distance to every other individual \cite{Freeman1979}.

The degree centrality of a node measures its potential communication activity or participation across the network, and is defined as the number of nodes adjacent to it \cite{Freeman1979}. The betweenness centrality of a node is defined as the proportion of the overall shortest paths between other pairs of nodes that pass through a particular node \cite{Freeman1979}. The betweenness centrality of node $n$ is computed as \cite{Freeman1979}:
\begin{eqnarray}
\label{formula:betweennessCentrality}
   betweenness(n)= \sum_{i,j} \frac{|p_{inj}|}{|p_{ij}|} %\nonumber
\end{eqnarray}
where, for each pair of nodes $(i,j)$ in the network, $p_{ij}$ are all the shortest paths linking $i$ and $j$, $p_{inj}$ are the ones that passing through node $n$. The intuition is that the larger the betweenness of a node becomes the more influence it will have on the information flow in the whole network, i.e., betweenness centrality is the index of the potential for control of communication \cite{Laniado2011,Freeman1979}. In the case of Wikipedia, nodes with higher  betweenness scores make the propagation of information and knowledge easier.

Eigenvector centrality is a measure of influence or popularity for nodes based on the adjacency matrix of the network \cite{Bonacich1987}, it takes into account a wide range of direct and indirect influences in the global network. The eigenvector centrality of individuals in the network is the eigenvector of the adjacency matrix that corresponds to the largest eigenvalue $\lambda$ of the adjacency matrix \cite{Bonacich1987}. Google's PageRank \cite{Page1998} measure is a variant of eigenvector centrality. We also make use of the PageRank variant to evaluate contributor authoritativeness for authors of Wikipedia articles.

%!TEX root = AICS2012-Wikipedia.tex
\section{Quality Measures}	%	Quality Scores or Quality Measures? Which is better?

% Explain how centrality and longevity measures are combined into scores.

% Sum(EditLongevity)

In this work, we employ three models to assess the quality of Wikipedia pages: an edit contribution-based model, a centrality-based model, and the combination of edit contribution and contributor authoritativeness. Our basic model is designed based on the principle that ``the higher the edit contribution of the article becomes, the better quality is the article''. This longevity-based model measures the quality of an article by aggregating the edit longevity of all its author contributions, and is defined as follows:  
\begin{eqnarray}
\label{formula:EditContribution}
   Longevity\_QScore(p) = \sum_{a \in A_{p} } contrib(a,p) %\nonumber
\end{eqnarray}

% sum(centrality)
Similarly, our centrality-based model is based on the principle that ``the higher the authority of the authors in a specific domain, the better quality is the article''. This model measures the quality of an article by the aggregation of authorities (i.e., centrality) from all its authors, and is defined as follows:  
\begin{eqnarray}
\label{formula:EditContribution}
   Cen\_QScore(p) = \sum_{a \in A_{p} } centrality(a) %\nonumber
\end{eqnarray}

While the longevity-based model captures the author contributions recorded in the edit history of the articles, the centrality based models mainly consider the contributor authoritativeness that encodes the communication patterns in the wikipedia networks. In contrast with the two previously mentioned models, a complicated way of assessing the quality of articles is to aggregate the edit longevity measure and contributor authoritativeness. The intuition for this is that each author plays a different role in the network (measured in terms of centrality), by nature some authors are more influential than others in the network. By incorporating this information in measuring author contribution to a page, we are expected to develop a better strategy to assess the quality of Wikipedia pages. For each author $a$ of page $p$, its contribution to $p$ can be computed as follows: 
\begin{eqnarray}
\label{formula:WeightedAuthorContribution}
   AuthorContrib(a,p) = contrib(a,p)\cdot centrality(a) %\nonumber
\end{eqnarray}
It is worth noting that the values of $contribute(a,p)$ and \\$centrality(a)$ for different authors can have different scales, so it is a good idea to normalize all these measures using minimum and maximum value from the whole dataset. Intuitively, a naive assessment of author contribution to a page would be the percentage of the final page contributed by each author (i.e., percentage normalization of edit contribution for each author to a page), weighted by the centrality of that author. However, this percentage normalization method neglects the information that high quality Wikipedia articles generally receive much more editions from Wikipedia editors, and leads to poor ranking of Wikipedia articles. Therefore, in the evaluation section, we choose to normalize $contribute(a,p)$ and $centrality(a)$ globally using min-max normalization.

To facilitate analysis in the following section, we denote the complex models derived from the combination of edit contribution with PageRank, eigenvector, degree, betweenness based centrality as $Com\_QScore_{PR}$, $Com\_QScore_{eigen}$,\\ $Com\_QScore_{degree}$ and $Com\_QScore_{btw}$, respectively. Then the quality score for page $p$ can be obtained by summing over the contributions of the set of its authors $A_{p}$ as follows:
\begin{eqnarray}
\label{formula:WeightedAuthorContribution}
   Com\_QScore(p) = \sum_{a \in A_{p} } AuthorContrib(a,p) %\nonumber
\end{eqnarray}

% Scale normalization:	x_new = (x_old-shift)/scale, shift = 0,   scale = max{abs(max), abs(min)}

%!TEX root = AICS2012-Wikipedia.tex
\section{Evaluation}

%  enwiki-20120211-pages-meta-history
%\vspace{-5.5 mm}
\begin{table*}[htb]
\centering
\caption{Statistics for the networks with $M$=10, $K$=20 and $\theta$=0.9} 
\label{tab:Statistics_network}
\begin{tabular}{|c|c|c|c|c|c|c|}

\hline
\multirow{2}{*}{Networks} & \multicolumn{3}{c|}{With bots} & \multicolumn{3}{c|}{Without bots}\\
\cline{2-7} 
 	& \#Authors & \#nodes  & \#edges & \#Authors & \#nodes  & \#edges \\
\hline
CoauthorNetwork & \multirow{3}{*}{18,844} & 18,844 & 628,524 & \multirow{3}{*}{19,606} & 19,606 & 712,685 \\
\cline{1-1}
\cline{3-4}
\cline{6-7}
TalkNetwork(signature) &  & 14,728 & 29,813 &  & 15,301 & 30,656 \\
\cline{1-1}
\cline{3-4}
\cline{6-7}
TalkNetwork(utpedits) &  & 17,034 & 704,248 &  & 17,700 & 723,088 \\
\hline

\end{tabular}
\end{table*}
%\vspace{-5.5 mm}

In this section, we provide an experimental evaluation of our approach on a dataset that consists of 9290 history-related articles from the WikiProject History\footnote{http://en.wikipedia.org/wiki/Wikipedia:WikiProject\_History} that fall under the FA (164), A (6), GA (312), B (906), C (900), Start (4072) and Stub (2930) classes. These articles have been assigned class labels according to Wikipedia Editorial Team’s quality grading scheme\footnote{More detail on the Wikipedia quality ratings can be found at \\http://en.wikipedia.org/wiki/Wikipedia:Version\_1.0\_Editorial\_Team/Assessment.}. The dataset was generated by extracting the complete edit history of the 9290 articles from the complete XML dump on $2012/02/11$\footnote{Available at http://dumps.wikimedia.org/enwiki/20120211/}. After collecting the dataset, we computed the edit contributions of all authors (i.e., edit longevity) using the WikiTrust software. Regarding author selection, we set the percentage contribution threshold $\theta$=0.9, minimum number of authors $K$=20, and minimum edit contribution $M$=10 \footnote{Varying the values of the parameters we did not observe remarkable differences in the experimental results.}. We generated two talk network versions for the WikiProject History: TalkNetwork(signature) and TalkNetwork(utpedits) by removing users who do not participate in the project from the two complete talk networks for the English Wikipedia. The difference between these two talk networks is explained at the end of section \ref{sec:related}. The co-author network for the project is also generated. The statistics for the networks are shown in Table \ref{tab:Statistics_network}. To get contributor authoritativeness, we performed centrality analysis on the talk networks and co-author network using Gephi software.

We employ the Normalized Discounted Cumulative Gain (NDCG) metric \cite{Jarvelin2002} to evaluate the performance of our quality measurement models to rank the Wikipedia articles. The NDCG metric was introduced by Jarvelin et al. \cite{Jarvelin2002} to measure the ability of a document retrieval algorithm to rank entries that are more relevant to the query. This metric has beed used by other researchers to evaluate the quality or trustworthiness of Wikipedia content \cite{Hu2007,Moturu2010} because it is suitable for ranking entries with multiple levels of assessment (e.g., Wikipedia quality ratings FA $\geq$ A $\geq$ GA $\geq$ B $\geq$ C $\geq$ Start $\geq$ Stub class labels). In the case of Wikipedia, we expect that articles that are more useful or more trustworthy to be ranked highly. NDCG is calculated as follows:
\begin{eqnarray}
\label{formula:NDCG}
   NDCG = \frac{1}{Z}\sum^{k}_{i=1} \frac{2^{s(r)}-1}{log(r+1)} %\nonumber
\end{eqnarray} 
where $Z$ is a normalization factor calculated so that a perfect ranking of the top $k$ articles would yield a NDCG of 1 and $s(r)$ denotes the score given to the article ranked at position $r$. In our case, we set different scores for different classes: $s(r)=6$ for featured article, $s(r)=5$ for A-star article % $s(r)=4$ for good article, $s(r)=3$ for B-class article, $s(r)=2$ for C-class article,
and so on down to  $s(r)=1$ for Start-class article, $s(r)=0$ for Stub-class article indicating that a Stub-class article at position $r$ does not contribute to the cumulative gain.

\subsection{Evaluation on All History-related Articles}
In this group of experiments, we evaluate our models using contributor authoritativeness derived from the three networks including and then excluding bot users. Table \ref{tab:NDCG_scores} presents the NDCG performance of our models on this dataset. In terms of excluding/including bots when assessing the quality, there is not much difference between the NDCG performance of the models on the three networks. It is apparent that the longevity-based model performs much better than centrality-based models in terms of NDCG score. 

%\vspace{-5.5 mm}
\begin{table*}[htbp]
\centering
\caption{NDCG scores for the whole Wikipedia History dataset} 
\label{tab:NDCG_scores}
\begin{tabular}{|c|c|c|c|c|c|c|}

\hline
\multirow{ 2}{*}{Models}	&\multicolumn{ 2}{c|}{CoauthorNetwork}&\multicolumn{2}{c|}{TalkNet(signature)}&\multicolumn{2}{c|}{TalkNet(utpedits)}\\
\cline{2-7}
		& w/o bots & with bots & w/o bots & with bots & w/o bots & with bots \\
\hline 
PageRank-based 		& 0.73 & 0.74 & 0.73 & 0.73 & 0.75 & 0.75 \\
\hline 
Eigenvector-based 	& 0.75 & 0.74 & 0.75 & 0.75 & 0.76 & 0.76 \\
\hline 
Degree-based 		& 0.74 & 0.74 & 0.75 & 0.75 & 0.76 & 0.76 \\
\hline 
Betweenness-based 	& 0.73 & 0.73 & 0.75 & 0.75 & 0.74 & 0.74 \\
\hline 
Longevity-based 	& 0.78 & 0.78 & 0.78 & 0.78 & 0.78 & 0.78 \\
\hline 
$Com\_QScore_{PR}$ 	& 0.77 & 0.77 & 0.77 & 0.77 & \textbf{0.79} & \textbf{0.79} \\
\hline 
$Com\_QScore_{eigen}$ 	& 0.77 & 0.77 & 0.78 & 0.78 & \textbf{0.79} & \textbf{0.79} \\
\hline 
$Com\_QScore_{degree}$ 	& 0.77 & 0.77 & 0.78 & 0.78 & 0.78 & 0.78 \\
\hline 
$Com\_QScore_{btw}$ 	& 0.76 & 0.77 & 0.78 & 0.78 & 0.78 & 0.78 \\
\hline

\end{tabular}
\end{table*}
%\vspace{-5.5 mm}

Regarding the complex models, on CoauthorNetwork and TalkNetwork(signature), we can see that their NDCG performance are very close to that of the longevity-based model; while on the TalkNetwork(utpedits), the complex models perform a little better than the longevity-based model (with an improvement about 1\% in $Com\_QScore_{PR}$ and \\$Com\_QScore_{eigen}$). It is also obvious that the NDCG performance of the models using contributor authoritativeness from TalkNetwork (utpedits) are slightly better than that of the corresonding models on CoauthorNetwork and TalkNetwork(signature). This indicates that the TalkNetwork(utpedits) is more informative (in terms of capturing the cummunications and interactions among authors) than other two networks. By inspecting on the talk network, we notice that the signature algorithm fails at extracting some active contributors (e.g., Borsoka\footnote{http://en.wikipedia.org/wiki/User\_talk:Borsoka}, Ptolemy Caesarion\footnote{http://en.wikipedia.org/wiki/User\_talk:Ptolemy\_Caesarion}, CarlosPn\footnote{http://en.wikipedia.org/wiki/User\_talk:CarlosPn}) when generating the network, which makes the Talk-Network (signature) smaller and less accurate. Overall, combining edit contribution and contributor authoritativeness to assess the quality of Wikipedia articles improves the NDCG performance to some extent, which implies that it is beneficial to take into consideration the contributor authoritativeness (i.e. communication patterns) when assessing the quality of Wikipedia content. One explanation for this may be that articles with significant contributions from authoritative contributors are likely to be of high quality, and that high-quality articles generally involve more communication and interaction between authors.

%\vspace{-5.5 mm}
%\begin{figure*}[!htb]
%\begin{center}
%  \subfloat{\includegraphics[width=0.48\textwidth]{./figs/PageRank+Stub-perCon-90-nobots-whole-dataset.eps}}\quad
%  \subfloat{\includegraphics[width=0.48\textwidth]{./figs/Betweenness+Stub-perCon-90-nobots-whole-dataset.eps}}\\
%  \subfloat{\includegraphics[width=0.48\textwidth]{./figs/EditCon+Stub-perCon-90-nobots-whole-dataset.eps}}\quad
%  \subfloat{\includegraphics[width=0.48\textwidth]{./figs/EditCon-PageRank+Stub-perCon-90-nobots-whole-dataset.eps}}
%\caption{Percentile distribution of wikipedia quality scores, using contributor authoritativeness from TalkNetwork(utpedits) without considering bot users.}
%\label{fig:PercentileDistribution}
%\end{center}
%\end{figure*} 
%\vspace{-5.5 mm}

\begin{figure*}[htb]
\begin{minipage}[b]{.48\linewidth}
  \centering
  \centerline{\epsfig{figure=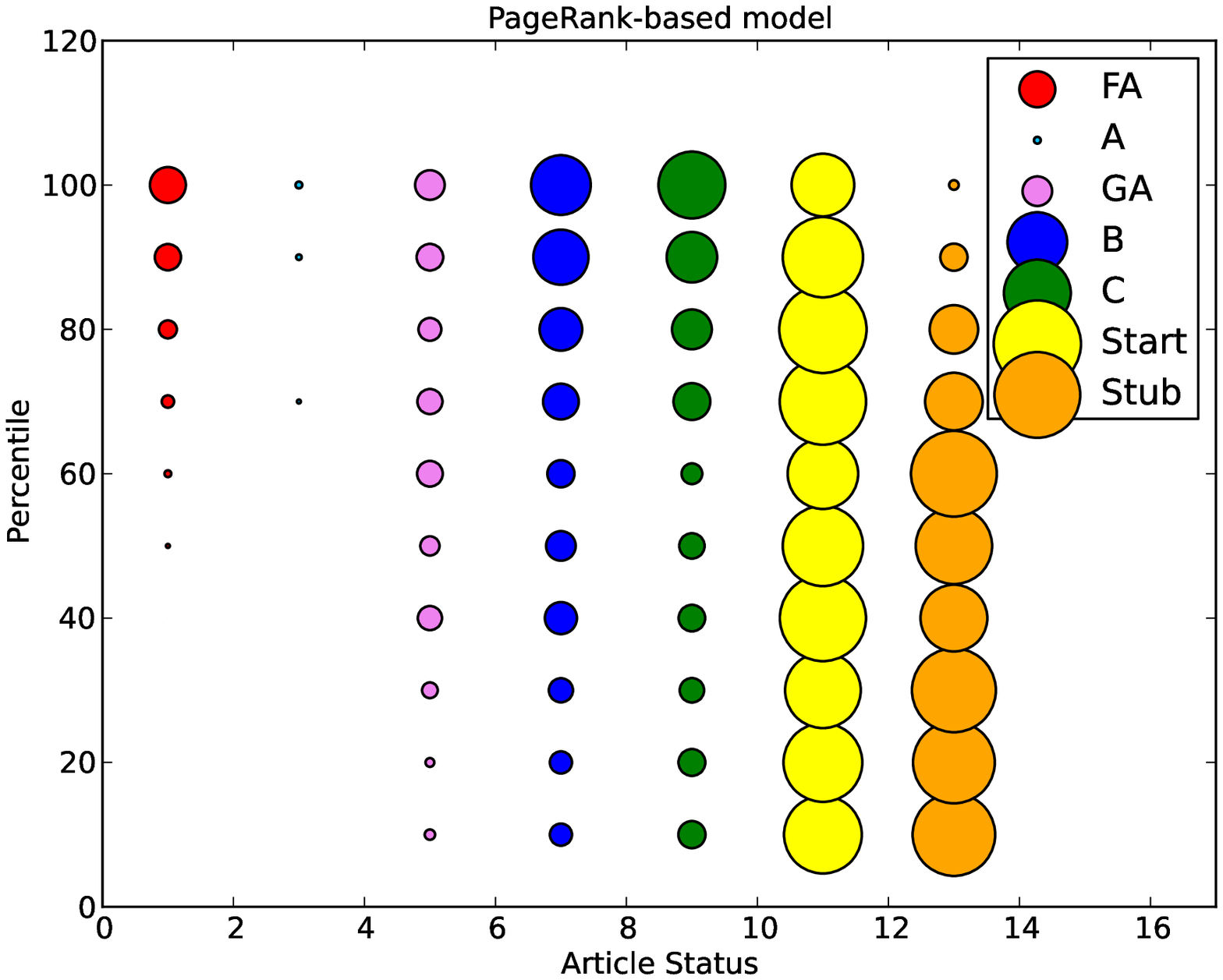,width=8.0cm}}
  %\vspace{1.5cm}
  %\centerline{(b) Results 2}\medskip
\end{minipage}
\hfill
\begin{minipage}[b]{0.48\linewidth}
  \centering
  \centerline{\epsfig{figure=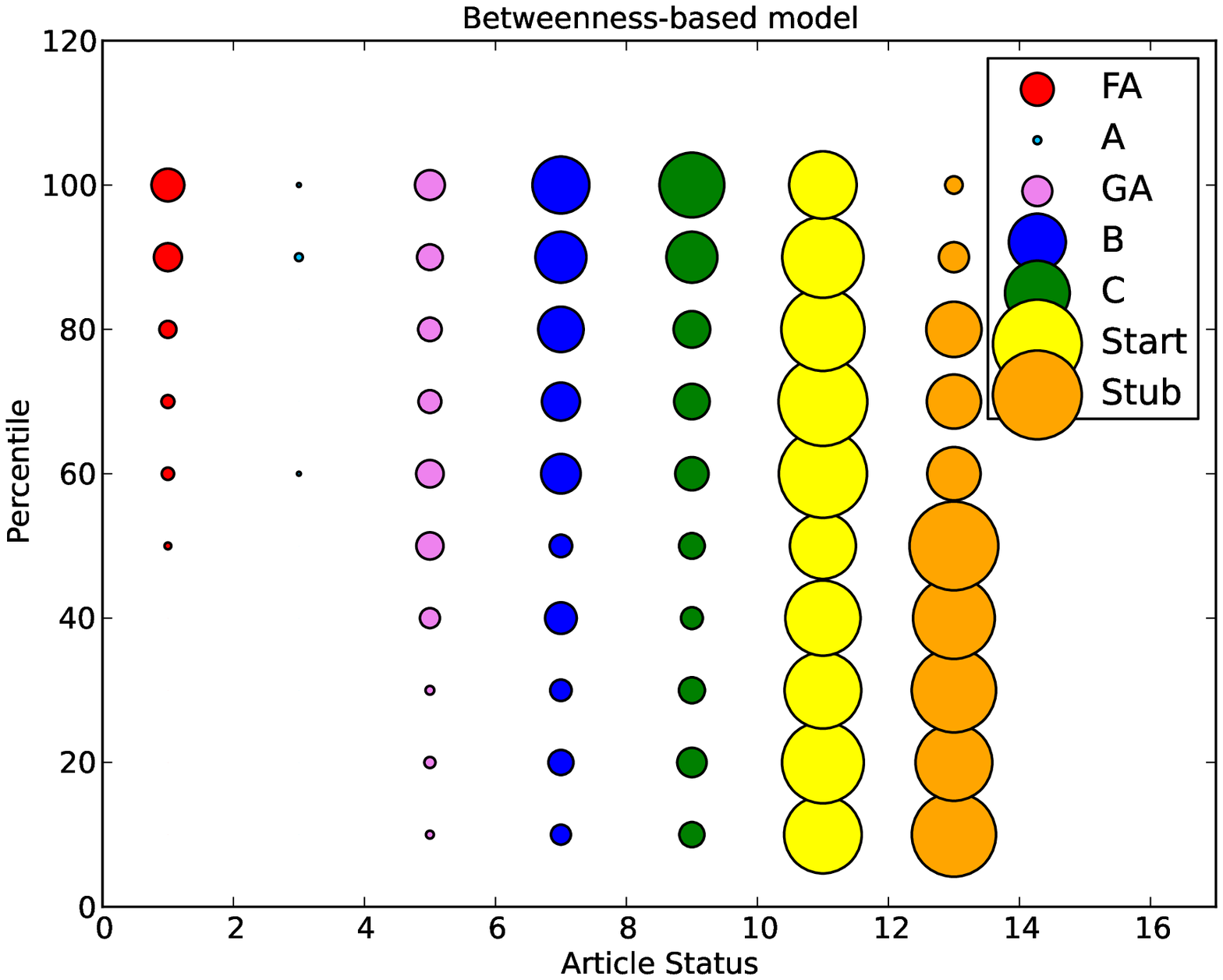,width=8.0cm}}
  %\vspace{1.5cm}
  %\centerline{(c) Result 3}\medskip
\end{minipage}
\begin{minipage}[b]{.48\linewidth}
  \centering
  %  EditCon+Stub-perCon-90-nobots-whole-dataset
  \centerline{\epsfig{figure=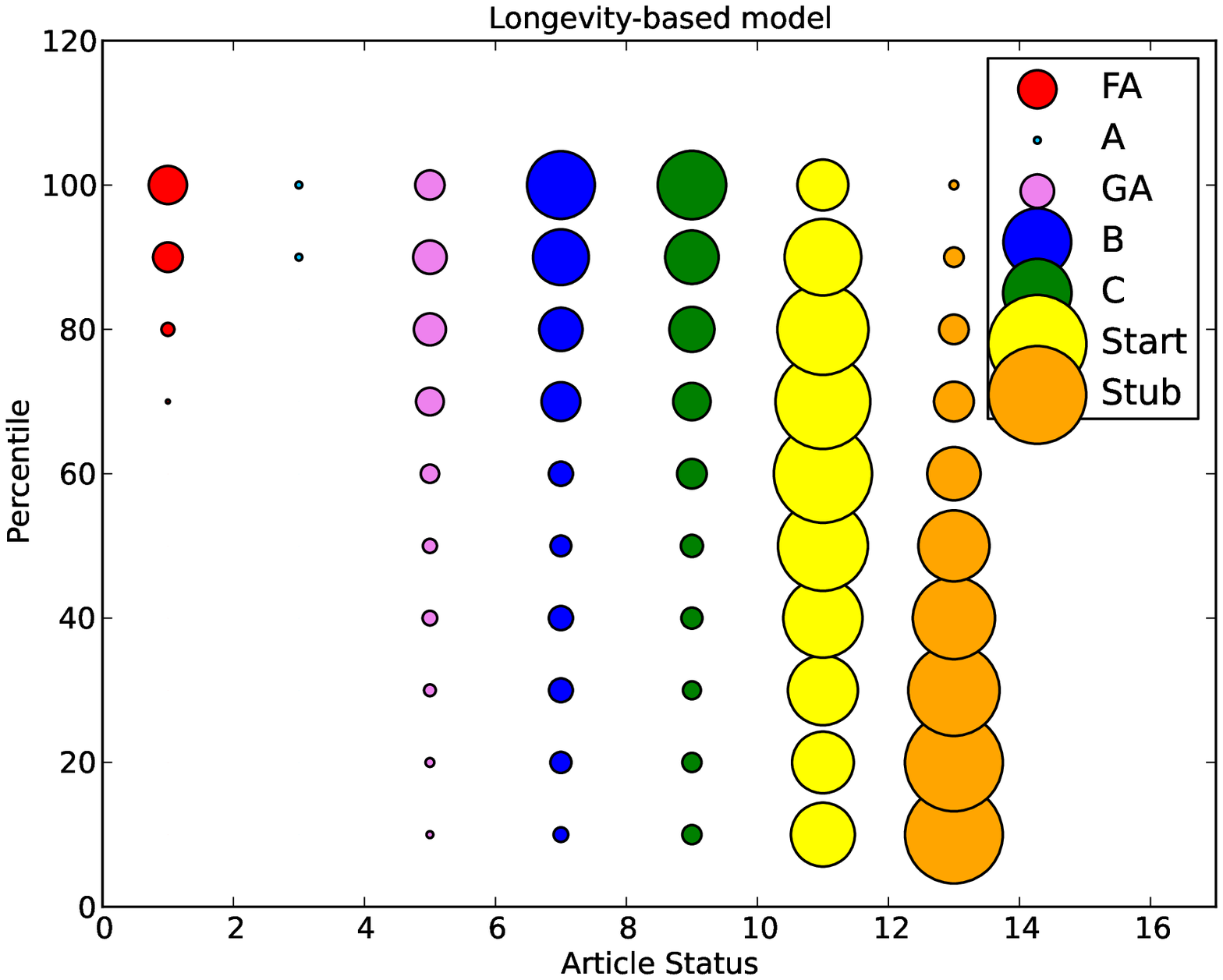,width=8.0cm}}
  %\vspace{1.5cm}
  %\centerline{(b) Results 2}\medskip
\end{minipage}
\hfill
\begin{minipage}[b]{0.48\linewidth}
  \centering
  \centerline{\epsfig{figure=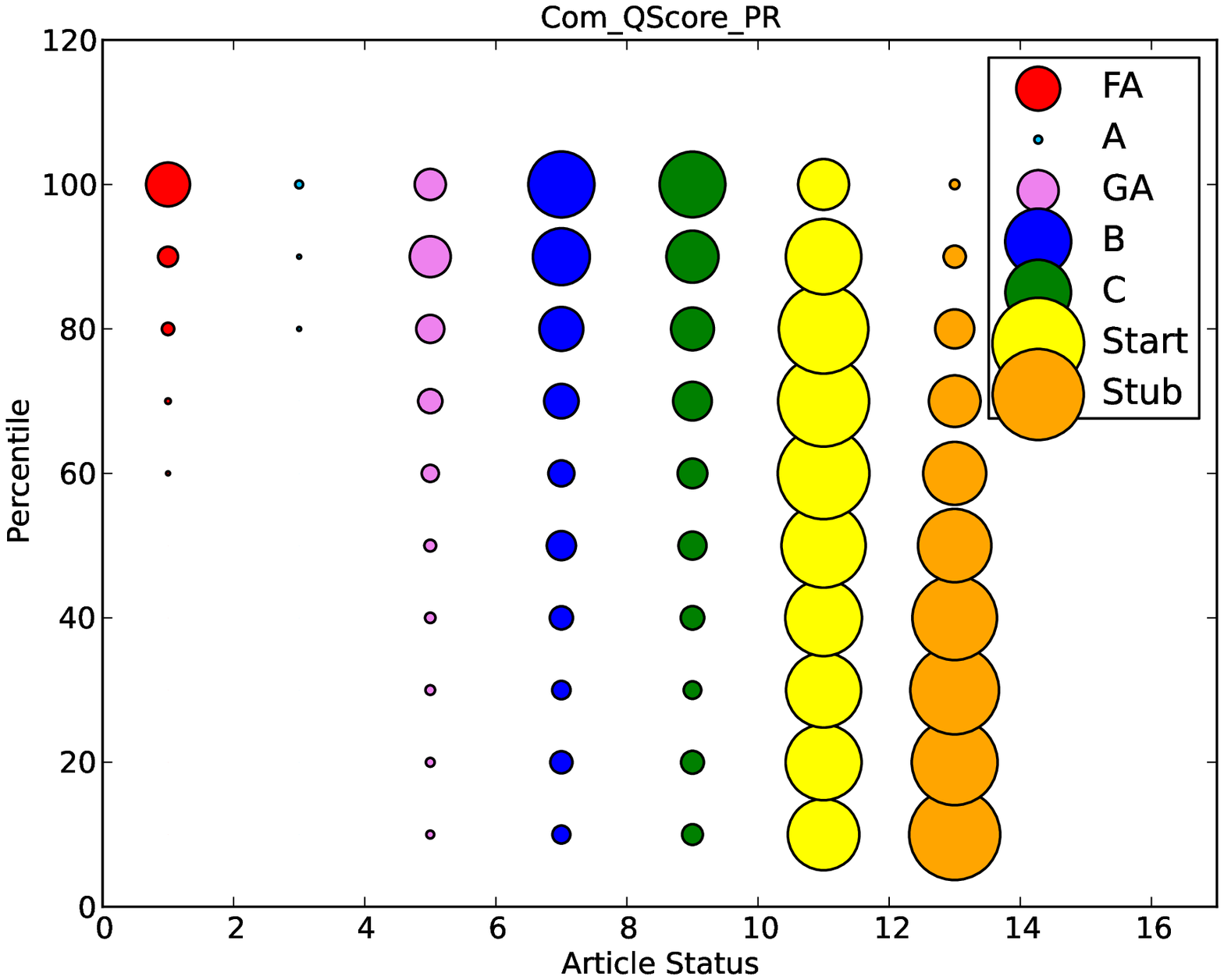,width=8.0cm}}
  %\vspace{1.5cm}
  %\centerline{(c) Result 3}\medskip
\end{minipage}

\caption{Percentile distribution of wikipedia quality scores, using contributor authoritativeness from TalkNetwork(utpedits) without considering bot users.}
\label{fig:PercentileDistribution}
\end{figure*}

To further understand the differences between these models, we also provide another view of the results from this experiment. In Fig. \ref{fig:PercentileDistribution} we plot the percentile distribution of Wikipedia quality scores (ordered by descending quality scores) for four models. The proportion of articles from a certain class that fall into a percentile is represented by a suitable sized bubble with color. We observe that there is a slight difference between the quality scores of the longevity-based model and complex model, with the latter ranking a bit more Featured articles in top 90 percentile than the former. All of the four models rank many Featured, A-class and Good articles very highly in top 90 percentile, the majority of Good, B and C-class articles are ranked dominantly in top 80 percentile, the largest amount of Start and Stub-class articles are ranked in the lower percentile. It is apparent that there is some overlapping between the classes, which indicates that no major difference exist between Featured and Good articles or between B and C-class articles. Overall, the percentile distribution of quality scores for the four models are very similar: with the quality scores for all the classes except Featured and A-class articles distributed across the whole percentile. 

It is worth noting that the quality scores of a small number of B, C, Start and even Stub-class articles are ranked very highly and appear in the top 90 percentile. By investigating those anomalously ranking articles, we find that: those highly ranking B or C-class articles are generally degraded from FA or GA status to its current status or are of good quality due to massive edits by Wikipedia users (e.g., History of classical music traditions\footnote{http://en.wikipedia.org/wiki/Talk:History\_of\_classical\_music\_traditions}, Stone Age\footnote{http://en.wikipedia.org/wiki/Stone\_Age}); those anomalous Start-class articles are mainly either degraded from other higher quality status (e.g., Monroe Doctrine\footnote{http://en.wikipedia.org/wiki/Monroe\_Doctrine} and Battle of Stirling Bridge\footnote{http://en.wikipedia.org/wiki/Battle\_of\_Stirling\_Bridge} degraded from FA status to Start-class) or are of relatively good quality and will be promoted to higher quality status as long as issues such as additional citations for verification being solved (e.g, at the time of writing the paper, History of Mexico\footnote{http://en.wikipedia.org/wiki/Talk:History\_of\_Mexico} was rerated from Start-class to C-class status, History of Scotland\footnote{http://en.wikipedia.org/wiki/History\_of\_Scotland} rerated from Start-class to B-class); the small number of highly ranked Stub articles are due to similar reasons such as being shortened from a longer version of the article which misused sources (e.g., Mathematics in medieval Islam\footnote{http://en.wikipedia.org/wiki/Mathematics\_in\_medieval\_Islam}). Basically, these highly ranked articles usually have some controversial issues which bring about edit war among the contributors. 

On the other hand, some Good, B and C-class articles are ranked in the lower 10 percentile. We find that these articles are rated as low importance by WikiProjects, and generally belong to short and concise articles that involve participation of very few authors (e.g., John of Argyll\footnote{http://en.wikipedia.org/wiki/John\_of\_Argyll}). Wu {\it et al.} \cite{Wu2011} also observe the similar trend when they visualize the network motif profiles of Wikipedia articles. To summerise, the percentile visualization of the quality scores provide some insights about analyzing the anomalous or outlier articles in the datasets, and can be used to help Wikipedia editors to identify Start and Stub articles that are of relatively good quality.

It is also worth noting that there are some differences between the percentile distribution of the quality scores we present here and the direct distribution of Wikipedia trust quality scores by Moturu and Liu \cite{Moturu2010} with regard to the type of distribution and the overlapping between different classes. Moturu and Liu plot the distribution according to the proportion of the normalized trust scores from a certain class that falls into each of the 11 Trust Score Categories (ranging from 0 to 10), we plot the percentile distribution of the scores based on the descending ranking of the quality scores. While our results contain some overlapping between Good, B, C and Start-class articles, their results contain less overlapping between different classes. We have explained the reasons for our anomalies above. A further reason for the difference may be that Moturu and Liu evaluate their models on a very small dataset of 230 health-related Wikipedia articles, while we evaluate our models on a larger dataset of 9290 history-related Wikipedia articles and we do not filter out the anomalous or controversial articles.  

\subsection{Evaluation on Filtering Dataset}

%\vspace{-5.5 mm}
\begin{table*}[!htb]
\centering
\caption{NDCG performance on filtering dataset with fewer classes.} 
\label{tab:NDCG_scores_filterDataset}
\begin{tabular}{|c|c|c|c|}

\hline
	 & Longevity-based & PageRank-based & $Com\_QScore_{PR}$\\
\hline
FA-C-Start-Stub & 0.806 & 0.778 & 0.837 \\
\hline
FA-C 		& 0.808 & 0.783 & 0.837 \\
\hline
FA-Start-Stub 	& 0.983 & 0.965 & 0.990 \\
\hline
FA-Start 	& 0.984 & 0.968 & 0.991 \\
\hline
FA-Stub 	& 0.995 & 0.992 & 0.996 \\
\hline 

\end{tabular}
\end{table*}
%\vspace{-5.5 mm}

% As other researchers also utilize limited classes of articles to evaluate their algorithms (e.g., \cite{Moturu2010}, \cite{Wu2011}), it is reasonable for us to evaluate our models using dataset with fewer classes. The motivation for this is that some of the classes of information are very close to each other, as we can see from Fig. \ref{fig:PercentileDistribution} about the overlapping of the distribution for GA, B, C and Start-class articles. To make a better comparison with existing studies that used limited clases, we evaluate our models on dataset with fewer number of classes and articles, using editor authorworthiness from TalkNetwork(utpedits) without considering bot users, as depicted in Table \ref{tab:NDCG_scores_filterDataset}. The distribution of this dataset is: FA(164), C(439), Start(414), Stub(280).

In the results presented in Fig. \ref{fig:PercentileDistribution}, it is not easy to get a clear picture of the relative performance of the different methods primarily because the information of some classes are very close to each other, which leads to the poor separation between these classes. In order to present a clearer picture we performed a further evaluation on some filtered datasets where there is clearer separation of the classes. This is consistent with evaluations performed by other researchers (e.g., \cite{Moturu2010}, \cite{Wu2011}). We use contributor authoritativeness from TalkNetwork(utpedits) without considering bot users, the result is depicted in Table \ref{tab:NDCG_scores_filterDataset}. The distribution of the filtered datasets is: FA (164), C (439), Start (413), Stub (279).

As expected, we observe an improvement in NDCG with the removal of A, Good, and B-class articles and the reduction of the number of C, Start and Stub-class articles. One reason for this is that with the elimination and reduction of these classes, which are very close to Featured and C, Start, Stub-class articles, there is clearer class separation. Moreover, the fact that removing Start and Stub articles shows little improvement in NDCG means that they are easy to distinguish from Featured and C-class articles--however there is clear overlapping between Featured and C-class articles. Finally, as with Moturu and Liu \cite{Moturu2010}, we also observe a maximum NDCG of 1 when only FA and Stub classes are considered. 

%\vspace{-5.5 mm}
%\begin{figure*}[!htb]
%\begin{center}
%    \includegraphics[width=0.75\textwidth]{./figs/FA-A-GA-Start-Stub-Precision-RecallCurve-standard-distribution-num_author-20.eps}
%\caption{Precision-recall curves on filtered dataset.}
%\label{fig:PrecisionRecallCurve}
%\end{center}
%\end{figure*}
%\vspace{-5.5 mm}

\begin{figure}[htb]

\begin{minipage}[b]{1.0\linewidth}
  \centering
  %  FA-A-GA-Start-Stub-Precision-RecallCurve-standard-distribution-num_author-20
  \centerline{\epsfig{figure=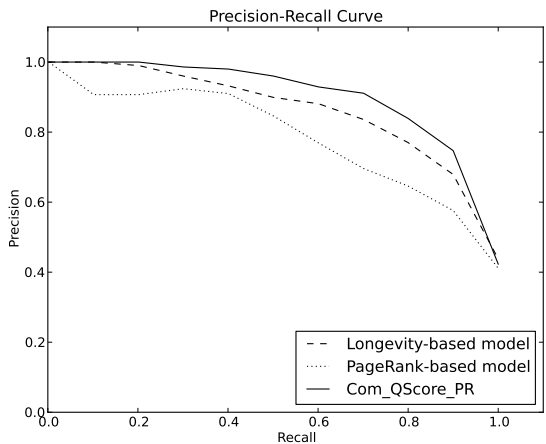,width=8.5cm}}
  %\vspace{2.0cm}
  %\centerline{(a) Result 1}\medskip
\end{minipage}

\caption{Precision-recall curves on filtered dataset.}
\label{fig:PrecisionRecallCurve}
\end{figure}

We also test on binary dataset which allows us to do some precision-recall analysis with which other researchers will be more familiar than NDCG analysis. The distribution of this dataset is: FA (164), A (6), GA (313), Start (414), Stub (280). We regard Featured, A-Class, Good articles as relevant to the target query, and remaining classes as irrelevant entries, the curve is shown in Fig. \ref{fig:PrecisionRecallCurve}. We observe from Fig. \ref{fig:PrecisionRecallCurve} that the curve closest to the upper right-hand corner of the graph is corresponding to the complex model, which indicates the best performance among the three models. The curve of the longevity-based model lies in the middle of the three curves, with the curve of PageRank-based model lying slightly below those of other two models. The precision-recall curves further prove that it is useful to take into account the inflence of users in the network of Wikipedia when assessing the quality of Wikipedia content.

\section{Conclusions and Future Work}\label{sec:conc}

In this paper, we study models for assessing the quality of Wikipedia pages based on edit contribution and contributor authoritativeness metrics. Edit contributions are quantified using edit longevity measure and contributor authoritativeness is scored using network centrality metrics in either the Wikipedia talk or co-author networks. We evaluate our quality measurement models on dataset that consists of 9290 history related articles from the WikiProject History.
 
The results suggest that it is useful to take into account the contributor authoritativeness (i.e., the centrality metrics of the contributors in the Wikipedia networks) when assessing the information quality of Wikipedia content. The implication for this is that articles with significant contributions from authoritative contributors are likely to be of high quality, and that high-quality articles generally involve more communication and interaction between contributors. The percentile visualization of the quality scores provides some insights about the outlier articles, and could be used to help Wikipedia editors to identify Start and Stub articles that are of relatively good quality. The proposed methods could also be used to suggest potential low quality articles which should be considered by experienced contributors.

At present, we do not have any special treatment to deal with reverted edits that do not introduce new content to a page. In some occasions, the edit longevity of users who reverted edits by malicious users to the last version dominate the edit contribution of that page. For instance, user HappyCamper\footnote{http://en.wikipedia.org/wiki/User\_talk:HappyCamper} made only one edit to Battle of Stirling Bridge\footnote{http://en.wikipedia.org/wiki/Battle\_of\_Stirling\_Bridge} by reverting the edit from an anonymous user (who rewrote the content of the whole page in a malicious way) to its last version. However, the edit longevity of this author to the page is 97995.8, which takes up about 92\% of the total edit contribution for the page. This is a misleading assessment of the author's contribution to that page and in turn distorts any resulting quality scores that use this assessment.

In this study, we impose the $|A_{p}|\geq K$ constraint on author selection so that we can select more authors even for those pages with their edit contributions dominated by a few contributors to compensate for this. In future work we will take steps to deal with reverted edits in the analysis as they can produce a false impression of author contribution. We noticed from our evaluation that there are some anomalous situations resulting from reorganization of the Wikipedia pages, for instance, an article may be considerably shortened by moving material to another page or a new page. Therefore, we should be more careful when using WikiTrust software to measure author contribution to Wikipedia articles, because in these cases the edit contribution obtained from the software may not be a real reflection of the author contribution. At present, we neglect the temporal information in the edit history of the Wikipedia pages and UTPs, it would be interesting to evaluate how the quality of the Wikipedia pages changes over time.

\section*{Acknowledgements}\label{sec:Acknowledge}

This work is supported by Science Foundation Ireland under Grant No. 08/SRC/I1407. Xiangju Qin is funded by University College Dublin and China Scholarship Council (UCD-CSC Scholarship Scheme 2011).

The statistical data used in this study to measure author contribution to Wikipedia articles are calculated by using the Wikitrust software. We would like to thank Luca De Alfaro, Bo Adler and Ian Pie for releasing the WikiTrust software. The talk networks used in this study are generated by using the wiki-network software, we also thank Paolo Massa for making this software available to the public.

\bibliographystyle{IEEEbib}
\bibliography{wiki-rep,strings,refs}

%\bibliographystyle{splncs}
%\bibliography{wiki-rep}

\end{document}